\documentstyle[12pt,epsf]{article}

\def \dfrac #1#2 {\displaystyle\frac{#1}{#2}}
\def\be{\begin{eqnarray}}
\def\ee{\end{eqnarray}}
\def\bq{\begin{equation}}
\def\eq{\end{equation}}
\def\ben{\begin{enumerate}}\def\een{\end{enumerate}}

\def\prl {Phys. Rev. Lett. }\def\pr{Phys. Rev. }
\def\np{Nucl. Phys. }

\def\roughly#1{\mathrel{\raise.3ex\hbox{$#1$\kern-.75em
\lower1ex\hbox{$\sim$}}}}

\begin{document}
\begin{titlepage}

\vspace{.3cm}
\hfill {\large FTUV 99-3; IFIC 99-3}
\vspace{.2cm}
\begin{center}
\ \\
{\LARGE \bf A Quark Model Analysis 
\\
\vspace{.1cm}
of Orbital Angular Momentum$\dagger$}
\ \\
\ \\
\vspace{.9cm}
{Sergio Scopetta and Vicente Vento$^{(a)}$}
\vskip 0.4cm
{\it Departament de Fisica Te\`orica}

{\it Universitat de Val\`encia}

{\it 46100 Burjassot (Val\`encia), Spain}
 
            and

{\it (a) Institut de F\'{\i}sica Corpuscular}

{\it Consejo Superior de Investigaciones Cient\'{\i}ficas}
\end{center}
\vskip 0.9cm
\centerline{\bf Abstract}
\vskip 0.4cm
Orbital Angular Momentum (OAM) twist-two parton distributions
are studied. At the low energy, {\sl hadronic}, scale we calculate them for the
relativistic MIT bag model and for non-relativistic potential quark models.
We reach the scale of the data by leading order evolution using the 
OPE and perturbative QCD.
We confirm that the contribution of quarks and gluons OAM
to the nucleon spin grows with $Q^2$,
and it can be relevant at the experimental scale,
even if it is negligible at the hadronic scale, irrespective of the model
used.
The sign and shape of the quark OAM distribution 
at high $Q^2$
may depend strongly on the relative size of the OAM and spin
distributions at the hadronic scale. 
Sizeable quark OAM distributions at the hadronic scale, 
as proposed by several authors, can produce 
the dominant contribution to the nucleon spin at high $Q^2$. 
As expected by general arguments,
we obtain,
that the large gluon OAM contribution is almost cancelled by the gluon spin 
contribution. 

\vskip 0.9cm
\leftline{Pacs: 12.39-x, 13.60.Hb, 13.88+e}
\leftline{Keywords:  hadrons, partons, orbital angular momentum, 
evolution.} 
\vspace{.6cm}

{\tt
\leftline {sergio.scopetta@uv.es}
\leftline{vicente.vento@uv.es}}

\vspace{0.6cm}
\noindent{\small$\dagger$Supported in part by DGICYT-PB97-1227 
and TMR programme of the European Commission ERB FMRX-CT96-008}.
\end{titlepage}

\section{Introduction}
\indent\indent

Understanding how the partons carry the angular momentum in 
the nucleon has become a main effort of present day physics. The quark spin 
contribution $\Delta \Sigma$ is well defined in QCD \cite{qspin}, 
thus measurable, and the gluon spin contribution $\Delta g$ is measurable,
despite some problems of definition \cite{gspin,ja-ma}, and
therefore both have been studied intensively. The situation with respect 
to the quark, 
$L_q$ and gluon, $L_g$, Orbital Angular Momentum (OAM) is less satisfactory.

It is well known, that the most natural definition of 
OAM for quarks and gluons cannot be separated in a gauge invariant way from
the corresponding spin terms \cite{ja-ma}. However, recently, new definitions 
of angular momentum have been implemented to accommodate gauge invariance
\cite{ji-2,ba-ja}, and from them adequate twist two OAM distributions 
have been constructed.

In this respect, three different possibilities have been investigated.
One proceeds by choosing a particular gauge \cite{ja-ma,ji-1, scha-1}. 
The OAM operator leads to the naive definition of orbital angular momentum 
up to effects not controlled by the gauge fixing. A second, 
maintains gauge invariance, by loosing covariance, 
defining the distributions only in the 
class of reference frame where the nucleon has a definite polarization 
\cite{hood}. In this case the resulting distributions can be related to
the forward limit of off-forward quantities, and are measurable  
\cite{ji-2,hood,hood-1}. The last proceeds by 
defining  OAM operators such that the distributions are
gauge invariant \cite{ba-ja} . Furthermore, in
the light-cone gauge, they reduce to the
natural definitions \cite{ja-ma}. At present, however, no physical
process has been proposed to access them.

Evolution equations for the OAM distributions have been derived
\cite{ji-1,scha-1,hood,har,ter}, and it has been shown \cite{hood}
that: 1) {\sl at} leading order (LO) the  evolution is determined by known
(polarized and unpolarized ) anomalous dimensions ;
2) {\sl at} LO the evolution equations for the different definitions
of the operators coincide; 
3) for the second procedure above, the evolution is related to that of 
the polarized and unpolarized distributions  {\sl at any} order. 

The  evolution equations
have been numerically  solved by using as input data-inspired
OAM distributions \cite{scha-2}.

During the last few years we have developed a scheme to study distributions
based on model calculations \cite{tr97}.
This procedure has been developed thus far only to
leading
order in the twist expansion and therefore we should compare with
the data only at high $Q^2$, where the contribution from the non
leading
twists vanishes.

 We have analyzed the quark and gluon
spin distributions, as well as odd chirality ones \cite{ss97}, in this scheme.
In this Letter we present the OAM distributions 
obtained by evolving those, 
 properly calculated within relativistic and non relativistic
models, to the experimental scale.

We present in the next section the theoretical framework behind our calculation
and discuss our definitions of the OAM distribution and their relation to the 
model distributions. The following section deals with the results of the 
evolution for several scenarios at the original hadronic scale. We analyze how 
hadronic structure influences the DIS results. Finally we extract in the 
conclusions the phenomenological implications of our study.

\section{The theoretical framework}

It has been suggested in the past that the OAM contribution to the nucleon spin
can be large at low \cite{vento,seg}, as well as large \cite{qspin}, energy 
scales. 
Since the quark spin, $\Delta \Sigma(Q^2)$, 
and the gluon spin, $\Delta g(Q^2)$, are 
observables, one may use the Spin Sum Rule
\cite{ja-ma} for the nucleon,

\bq
{1 \over 2} \Delta \Sigma(Q^2) + \Delta g (Q^2)
+ L_q(Q^2) + L_g(Q^2) = {1 \over 2}
\label{sr}
\eq
to determine the global OAM contribution to the spin. 
The problem of
separating 
the OAM in the quark and gluon fractions in a gauge invariant way,
already addressed in \cite{ja-ma}, is a cumbersome one.
In principle it has been solved in \cite{ba-ja}, where a
proper definition of the distribution functions for every term
in (\ref{sr}) has been introduced. Unfortunately these distributions
have not yet been related to observable quantities. 
Another approach to it has been proposed in Ref. \cite{ji-2},
where
the authors have shown 
that a gauge independent decomposition 
for the $total$ angular momentum, $J$, in quark, $J_q$, and gluon, 
$J_g$, contributions exists
in an interacting field theory. 
Moreover, $J_q$ and $J_g$ are gauge independent, and therefore they can be 
measured, for example, through a forward limit of off-forward 
quantities \footnote{A thorough discussion
of off-forward parton distributions (OFFPD) can be found in
 Refs. \cite{ji-3,rad}.}. Based on these developments, a gauge 
independent, twist-2, new definition for the quark OAM
distribution, whose evolution equation is known, has been given in ref.
\cite{hood}.  This definition
mixes the polarized and unpolarized singlet quark distributions
with the forward limit of OFFPD. Since the OAM matrix element
is not a Lorentz scalar, there is an ambiguity in this definition
for any relativistic quantum theory. To avoid it, one  must take for QCD 
a system of coordinates where the nucleon has a definite helicity, 
thus loosing covariance. 

We proceed to use this last  theoretical development, 
to perform a phenomenological model analysis of the OAM distributions, 
using relativistic as well as non-relativistic quark models. 

Let us first discuss the definition of the OAM for a relativistic model.
In this case we follow the definition provided by ref.
\cite{hood}, which is valid, 
in a class of coordinates in which the nucleon
has a definite helicity. If we  assume, for example, 
that the nucleon is moving in the $z$ direction and is polarized 
with helicity $+1/2$ , the
quark OAM distribution is given by,

\bq
L_q(x,Q^2) = {1 \over 2} \left[ x(\Sigma(x,Q^2) + E_q(x,Q^2)) -
\Delta \Sigma(x,Q^2) \right],
\label{lqr}
\eq
where $\Sigma(x,Q^2)$ $(\Delta \Sigma(x,Q^2))$ is the usual  
unpolarized (polarized) $singlet$ quark distribution
and $E_q(x,Q^2)$ is the $forward$ $limit$ of the
helicity-flip, chiral {odd}, twist-two OFFPD 
$E(x,\Delta^2,\Delta \cdot n)$ \cite{ji-2,ji-3}.
The latter quantity is defined through the twist-2 part
in the twist expansion for the following light cone correlation
function,

\begin{eqnarray}
\int {d \lambda \over 2 \pi} \langle P' |
\bar \psi(- \lambda n /2) \gamma^\mu \psi(\lambda n/2) |P \rangle
& = &  \nonumber \\
H(x, \Delta^2,\Delta \cdot n) \bar U(P') \gamma^\mu U(P)
& + & E(x, \Delta^2,\Delta \cdot n) \bar U(P') 
{i \sigma^{\mu\nu} \Delta_\nu \over 2 M} U(P)+ ...,
\label{corr}
\end{eqnarray}
where $P(P')$ is the 4-momentum of the initial (final) nucleon
in a virtual Compton scattering process, $\Delta_\mu=P'_\mu - P_\mu$,
$n = (1,0,0,-1)/2\Lambda$, $\Lambda$ is fixed 
by the choice of the reference frame, 
and $H(x, \Delta^2,\Delta \cdot n)$ is the helicity-conserving,
chiral even, twist-2
OFFPD whose forward limit is the usual forward
unpolarized parton distribution.
In the definitions above, the forward limit corresponds to
$\Delta^2 \rightarrow 0$, $\Delta \cdot n \rightarrow 0$.

In the non-relativistic case, the nucleon
wave function is given in general by an expansion in terms of the 
eigenstates of
some approximate hamiltonian. Let N label the quantum numbers of the
eigenstates, i.e., principal quantum number, 
orbital angular momentum, spin, ..., then

\bq
\Psi(\vec p_1, \vec p_2, \vec p_3) = 
\sum_{N} a_{N} 
\psi_{N} (\vec p_1,\vec p_2,\vec p_3)
\label{calL}
\eq
The OAM quark parton distribution, generalizing our
approach developed for unpolarized and polarized distributions
\cite{tr97}, is determined by,

\bq
L_q(x,Q^2) = 2 \pi M \sum_{N} 
|a_{N}|^2 \int_{|p_-(x)|}^\infty 
\, dp \, p \, L_{q,{N}}^z(p)~,
\label{lqnr}
\eq
where

\bq
L_{q,{N}}^{z}(p) = 
\langle \psi_{N} (\vec p_1,\vec p_2,\vec p_3)|
\sum_{i=1}^3 L_i^{z} \delta (\vec p - \vec p_i)
|\psi_{N} (\vec p_1,\vec p_2,\vec p_3) \rangle,
\label{lqz}
\eq
and the lower integration limit is 

\bq
|p_-(x)| = {M \over 2} \left [ x -\left( m_q \over M \right )^2
{1 \over x} \right ]~.
\label{pm}
\eq

Having set down the framework which defines the OAM distributions 
at the hadronic scale, we proceed to calculate them explicitly 
in two models and
to study their evolution.

\section{Analysis of Orbital Angular Momentum distributions at high $Q^2$}

We proceed to study the OAM distribution in two different scenarios for proton
structure: i) a non-relativistic scheme based on the Isgur-Karl model 
\cite{ik78};
ii) a relativistic approach, as described by the MIT bag model \cite{ja74}.

For the non-relativistic scenario of proton structure we consider initially
the Isgur-Karl model with a proton wave function given by a harmonic oscillator
potential 
including contributions up to the $2 \hbar \omega$ shell
\cite{gia}. In this case  the wave function, Eq. (\ref{calL}), is given by the
following admixture of states

\bq
|N \rangle = 
a_{\cal S} | ^2 S_{1/2} \rangle_S +
a_{\cal S'} | ^2 S'_{1/2} \rangle_{S} +
a_{\cal M} | ^2 S_{1/2} \rangle_M +
a_{\cal D} | ^4 D_{1/2} \rangle_M~,
\label{ikwf}
\eq
where we have used the spectroscopic notation $|^{2S+1}X_J \rangle_t$, 
with $t=A,M,S$ being the symmetry type.
The coefficients were determined by spectroscopic properties to be: 
$a_{\cal S} = 0.931$, 
$a_{\cal S'} = -0.274$,
$a_{\cal M} = -0.233$, $a_{\cal D} = -0.067$.

Calculating Eq. (\ref{lqz}) using the wave function Eq. (\ref{ikwf}),
one gets for the parton distribution, Eq. (\ref{lqnr}),

\bq
L_q(x,\mu_0^2) = |a_{\cal D}|^2 { M \over \alpha \sqrt{\pi}}
\left ( {3 \over 2} \right )^{3/2} 
\left( 
{1 \over 5} {p_-^4 \over \alpha^4} + 
{13 \over 30} {p_-^2 \over \alpha^2} + 
{23 \over 45} 
\right) 
e^{ - { 3 p_-^2 \over 2 \alpha^2} }~,
\label{lqik}
\eq
where $\alpha^2$ is a  parameter of the model,
and $p_-(x)$ is given by (\ref{pm}).
Note that only the small $|^4 D_{1/2}\rangle_M$ 
wave component gives a contribution to this OAM distribution.

In the relativistic scenario, we proceed to evaluate in 
the MIT bag model 
the twist-two distribution Eq. (\ref{lqr}).
The term $\Delta \Sigma(x,\mu_0^2)$ in Eq. (\ref{lqr})
is discussed for the bag
in ref. \cite{ja-ji}.

To calculate Eq.(\ref{lqr}) we need the quantity
$\Sigma(x,\mu_0^2)+E_q(x,\mu_0^2)$, which is given by  the forward
limit, 
$\Delta^2 \rightarrow 0$, $\Delta \cdot n \rightarrow 0$, 
of the OFFPD distribution 
$H(x,\Delta^2,\Delta \cdot n)+
E(x,\Delta^2,\Delta \cdot n)$. The result for the latter in the bag, 
calculated by  using  the general definition (Eq. (\ref{corr})), is
to be found in Eq.(29) of ref. \cite{meln}, 
from which we have obtained 
the forward limit, i.e.,

\begin{eqnarray}
\Sigma(x,\mu_0^2) +  E_q(x,\mu_0^2)  & = & 
Z^2(0) {N^2 R^6 M \over 2 \pi^2} 
\int \, d k_{\perp}  d \phi 
\, k_{\perp} \, \{  \, t_0^2(k) + 
\\ \nonumber
& + & 2 M \eta { t_0(k)t_1(k) \over k } + 
2M {k_x \over k k'}
\left ( { k t_0(k) t_1(k') - k' t_1(k) t_0(k') \over \Delta_x} \right ) +  
\\ \nonumber
& + & M^2 \left ( x - { 1 \over 4} \right )
\left (2 \eta -  
x + { 1 \over 4} \right ) {t_1^2(k) \over k^2} \, \}
\end{eqnarray}

where
\bq
N^2 = {\omega_0 \over 2 R^3 (\omega_0-1)j_0^2(\omega_0)}
\eq
and
\bq
Z(0) = N^2
\int_0^R \, dr \, r^2 \, \left ( j_0^2 \left ( {M \over 4} r \right )
+ j_1^2 \left ( {M \over 4} r \right ) \right )
\eq
$\omega_0=2.04$ is the lowest energy eigenvalue for the quark in the bag, 
$j_{0(1)}$ are the spherical Bessel functions,
$k$ is the momentum of the  quark in the bag,
$k'=|\vec k´|, \, \vec k' = \vec k + \eta \vec \Delta$, 
$\eta$ is a parameter of the calculation in \cite{meln}
and, eventually
\bq
t_0(k) = {j_0(\omega_0) \cos(kR) - j_0(kR) \cos \omega_0
\over \omega_0^2 - \vec{k}^2 R^2}
\eq
and
\bq
t_1(k) = {j_0(kR) j_1(\omega_0) \cos(kR) - 
j_0(\omega_0) j_1(kR) \omega_0
\over \omega_0^2 - \vec{k}^2 R^2}.
\eq

We recall, that the calculation of 
$H(x,\Delta^2,\Delta \cdot n)+
E(x,\Delta^2,\Delta \cdot n)$
in \cite{meln}, whose forward limit we are using here,
is performed in the Breit reference frame. 
The bag model has no translational invariance and  therefore the choice
of frame must be associated to the model  assumptions.
 Our scheme requires that the evaluation is performed in the rest
frame, thus 
we have to fix here the parameter $\eta$ to the value of 0.75
which corresponds, according to Ref. \cite{meln}, to the
{\sl unboosted} calculation.

In both cases, Isgur-Karl (IK) and MIT,
we use the corresponding support correction 
as defined in \cite{tr97} and \cite{jr80}, respectively.

Once the distributions $L_q(x,\mu_0^2)$
have been obtained at the low (hadronic) scale 
of the model, we perform a LO QCD evolution according 
to the equations displayed in 
refs. \cite{scha-1,hood,har,ter} and numerically solved in \cite{scha-2}.
These contain a complicate mixing between 
$L_{q(g)}(x,\mu_0^2)$, $\Delta \Sigma(x,\mu_0^2)$
and $\Delta g(x,\mu_0^2)$, a feature 
which will be very relevant in the analysis
of the data.

The results of our analysis are shown in Figs. 1 through 5.

In Fig. 1 we show the  IK result for quarks (gluons) in (a) ((b));
The full curve corresponds to the  initial distribution, which is missing 
in (b), since we start from a very low hadronic scale where no valence 
gluons exist. At LO the hadronic scale corresponds to $\mu_0^2\simeq0.08$ 
$GeV^2$. The dashed curves correspond to the result of the evolution from the
hadronic OAM distributions, to $Q^2=10$ $GeV^2$ (short-dashed) and to
 $Q^2=1000$ $GeV^2$ (long-dashed).

We can summarize the results of the calculation as follows:
i) the evolved distributions are negative;
ii) the magnitude of the distributions increases with $Q^2$
at low $x$; iii) the magnitude of $L_q(x,Q^2)$ is {\sl small} but increases
with respect to the tiny starting distribution
(Eq. (\ref{lqik})); iv) though gluons are assumed
to be negligible at the hadronic scale, $L_g(x,Q^2)$ becomes much larger
than $L_q(x,Q^2)$ at high $Q^2$.

We show in Fig. 2 the same analysis for the MIT bag model.
Here the initial $L_q$ is much larger, but the result of the evolution is
 qualitatively basically the same.

Thus our first conclusion is that there is little model dependence for 
 different
initial OAM distributions. In this case their structure does not seem to 
influence very much the evolution. It is clear that the other inputs of the 
equations, $\Delta \Sigma (x,\mu_0^2)$ and $\Delta  g (x,\mu_0^2)$,
are the dominating features.

For these results the initial scale has been very low and therefore
no initial gluon distribution was required.
In Fig. 3 we show the result of the evolution for the IK model
starting from a higher scale, $\mu_0^2\simeq 0.23$ $GeV^2$.
At this hadronic scale, for LO, about 40 $\%$ of the proton  momentum must be 
carried by the gluons.
The polarized gluon contribution is built starting from
the valence quark distributions as done in \cite{tr97}
and suggested in \cite{reya}, and we define $L_g(x,\mu_0^2)$ from
$L_q(x,\mu_0^2)$ using the same prescription.

Again the same
features as in the first analysis are found. The initial
gluon distribution is so small, in relation with the final one, that
it does not show up in (b).

We note that this choice of the initial scale is the same adopted 
in the numerical analysis of ref. \cite{scha-2}.
An important check of the consistency in our evolution is that 
we have recovered the  results of this reference by inserting their 
initial distributions in our code.

Recapitulating, our results show that the input distributions 
$L_q(x,\mu_0^2)$ and $L_g(x,\mu_0^2)$
do not seem to determine the behavior of their evolved 
ones, which turns out to be governed by the 
the singlet
polarized distributions $\Delta g (x,\mu_0^2)$ and 
$\Delta \Sigma(x,\mu_0^2)$, due to their mixing in the evolution 
equations. To check to what extent such a statement, stressed also in
\cite{scha-2}, is valid,
we analyze in Fig. 4 the results of 
a modification of the IK model, the so called  D model,
already studied in \cite{ropele}.
In this variant model,
the $D$ wave probability is set large to reproduce
the axial coupling constant of the nucleon \cite{vento}. This condition
requires the following choice
of the parameters in Eq. (\ref{ikwf}):
$a_{\cal S} = 0.894$, $a_{\cal S'} = 0$,
$a_{\cal M} = 0$, $a_{\cal D} = -0.447$,
i.e., the probability to find a nucleon in the $D$ wave
is about 20 $\%$.  Moreover we have introduced also in the D Model
scenario polarized valence gluons, as we did for the IK scenario of Fig.3.
From the figure it is clear that, while the result for the gluons 
does not differ in a relevant way from the ones found before 
with the various models, the result for $L_q(x,Q^2)$ does, this distribution 
becoming rather large and positive for large $x$. It is important to
stress that in order for this to occur two mechanisms were needed, a
large initial OAM distribution (as in the MIT bag Model), and a higher
hadronic scale (as provided by the valence gluons). As shown
previously, the independent action of the two mechanisms does not lead
to this behavior. It is the joint action of large initial OAM
distributions and a softer evolution which is responsible for it.  
Thus, $\Delta g (x,\mu_0^2)$ and 
$\Delta \Sigma(x,\mu_0^2)$ are governing
the evolution  as long as they are much larger than
$L_q(x,\mu_0^2)$ at the initial scale. When they have similar size
the above statement is not true any more.
Note that the IK interaction together
with the choice of parameters of the  D model does not
describe the hadron spectrum. Nonetheless, other models
of interaction (for example \cite{vento,seg}) predict at the low scale 
a large OAM and fit the spectrum. 
We conclude therefore that a precise knowledge of the OAM distributions 
will serve to distinguish among the models.

In Fig. 5 we show the evolution of the various 
contributions to the spin sum rule, Eq. (\ref{sr}).
We write here below the LO-evolution equations for the first
moments of the distributions involved, corresponding to
three active flavors (cf. \cite{scha-2}):
\begin{eqnarray}
\Delta \Sigma (Q^2) & = & \Delta \Sigma(\mu_0^2)~, \\
\Delta g (Q^2) & = & {4 \over 9} 
\left ( 
b^{-1}
- 1 
\right ) 
\Delta \Sigma(\mu_0^2) + 
b^{-1}
\Delta g(\mu_0^2)~, \\
L_q (Q^2) & = &  
\left ( 
b^{-{38 \over 81}}
- 1 
\right ) 
{1 \over 2} \Delta \Sigma(\mu_0^2) + 
b^{-{38 \over 81}}
L_q(\mu_0^2) - 
{9 \over 50} 
\left ( 
b^{-{38 \over 81}}
- 1 
\right )~, \\
L_g (Q^2) & = &  
b^{-{38 \over 81}}
\Delta g (\mu_0^2) 
- \Delta g (Q^2)
+ b^{-{38 \over 81}}
L_g (\mu_0^2)
- {8 \over 25} \left ( 
b^{-{38 \over 81}}
- 1 
\right )~,
\end{eqnarray}
where $b = {\alpha_s(Q^2)/ \alpha_s(\mu_0^2) }$.

Fig. 5 (a) corresponds to the modified IK scenario
used in Fig. 3, where at the scale of the model
the OAM carried by quarks 
and gluons is very small and the rest of the proton spin
is almost equally shared between quarks and gluons spins
($L_q(\mu_0^2) + L_g(\mu_0^2)\simeq 0.01$,
$\Delta \Sigma(\mu_0^2)\simeq 0.48 $ and
$\Delta g(\mu_0^2)\simeq 0.25 $),
whereas Fig. 5 (b) corresponds to the extreme
scenario used already in Fig 4, the so called  D model,
with a large initial OAM
($L_q(\mu_0^2)=0.145$, $L_g(\mu_0^2)= 0.055$,
$\Delta \Sigma(\mu_0^2)\simeq 0.4 $ and
$\Delta g(\mu_0^2)\simeq 0.1 $).  As predicted by total angular 
momentum conservation \cite{qspin} and already obtained
in \cite{scha-2} as a model-independent feature
of the evolution equation , it is seen that at large $Q^2$
the huge negative contribution $L_g(Q^2)$ basically
cancels out with the positive $\Delta g(Q^2)$.
Anyway, the role of the quark OAM is found to be very
important in the second scenario (cf. Fig. 5(b)),
being at large $Q^2$ the largest contribution to the saturation
of the spin sum rule.
Again, we see that quark OAM, due to evolution, can
be important at large $Q^2$ if it is not negligible at the
scale of the model, whereas the gluons OAM, though
it is large, is basically cancelled by the gluons spin. 

 Before concluding this section a comment on our perturbative evolution 
is advisable. As has been discussed previously the scale of the models
is rather low. It is therefore natural to think that NLO corrections are
relevant and could change the interpretation of the results.
Actually, our past experience \cite{tr97,ss97} in similar situations
suggests that NLO corrections are at the level $30-40 \%$ at most, 
and moreover, they never spoil the LO behavior.
We are thus confident that the main results of our analysis
will survive, at least in a qualitative fashion, after the implementation 
of NLO corrections.

\section{Conclusions}

We have studied the OAM distributions as defined newly in order to take into
account gauge invariance. Our study, as in previous occasions \cite{tr97}, 
is at
the heart of the ill-labelled spin crisis. We define a hadronic scale at which
quark models should be operative. We use the quark model calculations as
non-perturbative input for the evolution equations and therefore to predict the
experimental outcome at the DIS scale. We have seen in this way that much of
the folklore, associated with the spin crisis, has banished. However the 
procedure has served to distinguish, in detailed calculations, among the 
different models. In here we perform an analysis of the OAM 
distributions.

We have seen that evolution, as in the other calculations, plays a major role 
in the outcome of the predictions. The fact that the gluon
and quark spin singlet
distributions mix in these equations with the quark
OAM distributions, implies that
for large $Q^2$ large contributions from the OAM are to be expected, even if
they are not present at the hadronic scale. Thus, two scenarios arise in a
natural way. One, the more conventional one, as described by the more
traditional models, is defined by quark OAM distributions at a small
hadronic scale.
In this scenario the evolved distributions are large and negative and 
almost model
independent and angular momentum DIS physics is dominated by the quark and
gluon spin singlet distributions, 
not by OAM distributions at the hadronic scale. 
A second scenario is defined by quark and gluon OAM distributions at a higher 
hadronic scale. In the latter, soft evolution scenario,
the initial
distributions are important and therefore DIS 
physics may be able to discriminate between different models. If the
OAM distributions are large the outcome of the evolution is very
strongly dependent on the initial distributions and completely
different from that of the first scenario.

Finally the gluon OAM distributions become huge through evolution, 
even if they are
not present at the hadronic scale. 
However, as it is well known 
\cite{qspin} , 
the gluon OAM and gluon spin
contributions cancel to a great extent in the nucleon spin, 
but not so in other
moments.

 Our past experience suggests that LO results provide a 
reasonable {\sl qualitative} 
approximation and we do not expect that NLO corrections
 can spoil their general features.

It is clear that many phenomenological implications have arisen of our study.
A careful analysis of gauge invariance \cite{ji-2,ba-ja,hood} 
has permitted us to obtain
many observables, which may not only lead to a better understanding 
of the proton
spin, but to describe the proper behavior of QCD at low energies, i.e.,  in the
confining region.
These observations, implemented in our scheme, are instrumental in defining 
the picture of the proton at the hadronic scale that should be used for
describing low energy properties.
  
\section*{Acknowledgments}

We thank S.V. Bashinsky and R.L. Jaffe for valuable comments and 
suggestions, and X. Ji, W. Melnitchouk and X. Song for 
clarifications on their calculation of Ref. \cite{meln}.

\newpage
\centerline{\bf \Large Captions}
\hfill\break
\hfill\break
\hfill\break
{\bf Figure 1}: 
Proton OAM distributions in the IK model for the quarks, (a), and for the
gluons, (b).
The full curves show the initial distributions
at the hadronic scale of $\mu_0^2=0.08$ $GeV^2$, where
a negligible fraction of the nucleon momentum is carried by the gluons; 
the dashed curves represent the LO--evolved distributions
at $Q^2=10$ $GeV^2$; the long-dashed curves give the
LO--evolved distributions at $Q^2=1000$ $GeV^2$.
\hfill\break
\hfill\break
{\bf Figure 2}: 
The same as in Fig.1, but for the MIT model.
It should be noticed that the distribution 
at the hadronic scale of $\mu_0^2=0.08$ $GeV^2$
is shown as it comes
out from the model calculation, before the support correction is
implemented.
For this reason it does not go to zero at $x=1$.
\hfill\break
\hfill\break
{\bf Figure 3}: 
Proton OAM distributions in a modified IK model (see text)
for the quarks, (a), and the gluons, (b).
The full curves show the initial distributions
at the hadronic scale of $\mu_0^2=0.23$ $GeV^2$, where
around 40 \% of the nucleon momentum is carried by the gluons;
the dashed curves represent the LO--evolved distributions
at $Q^2=10$ $GeV^2$; the long-dashed curves give the 
LO--evolved distributions
at $Q^2=1000$ $GeV^2$.
\hfill\break
\hfill\break
{\bf Figure 4}: 
Proton OAM distributions for the ``D-Model'' (see text),
for the quarks, (a), and the gluons, (b).
The full curves show the initial distributions
at the hadronic scale of $\mu_0^2=0.23$ $GeV^2$, where
around 40 \% of the nucleon momentum is carried by the gluons;
the dashed curves represent the LO--evolved distributions
at $Q^2=10$ $GeV^2$; the long-dashed curves give the
LO--evolved distributions at $Q^2=1000$ $GeV^2$.
\hfill\break
\hfill\break
{\bf Figure 5}: The contributions to the proton spin sum rule,
Eq. (\ref{sr}), according to: (a) the modified IK scenario
of Fig. 3; (b) the ``D model'' scenario of 
Fig. 4. The dashed curve shows ${1 \over 2}\Delta \Sigma(Q^2)$,
the long-dashed one $\Delta g (Q^2)$, the dot-dashed curve is $L_q(Q^2)$,
the dot-long-dashed curve gives $L_g(Q^2)$ and the full curve 
represents the sum of the 
previous four terms, giving the spin sum rule
($J={1 \over 2}$).

\newpage
\begin{figure}[h]
\vspace{10cm}
\includegraphics{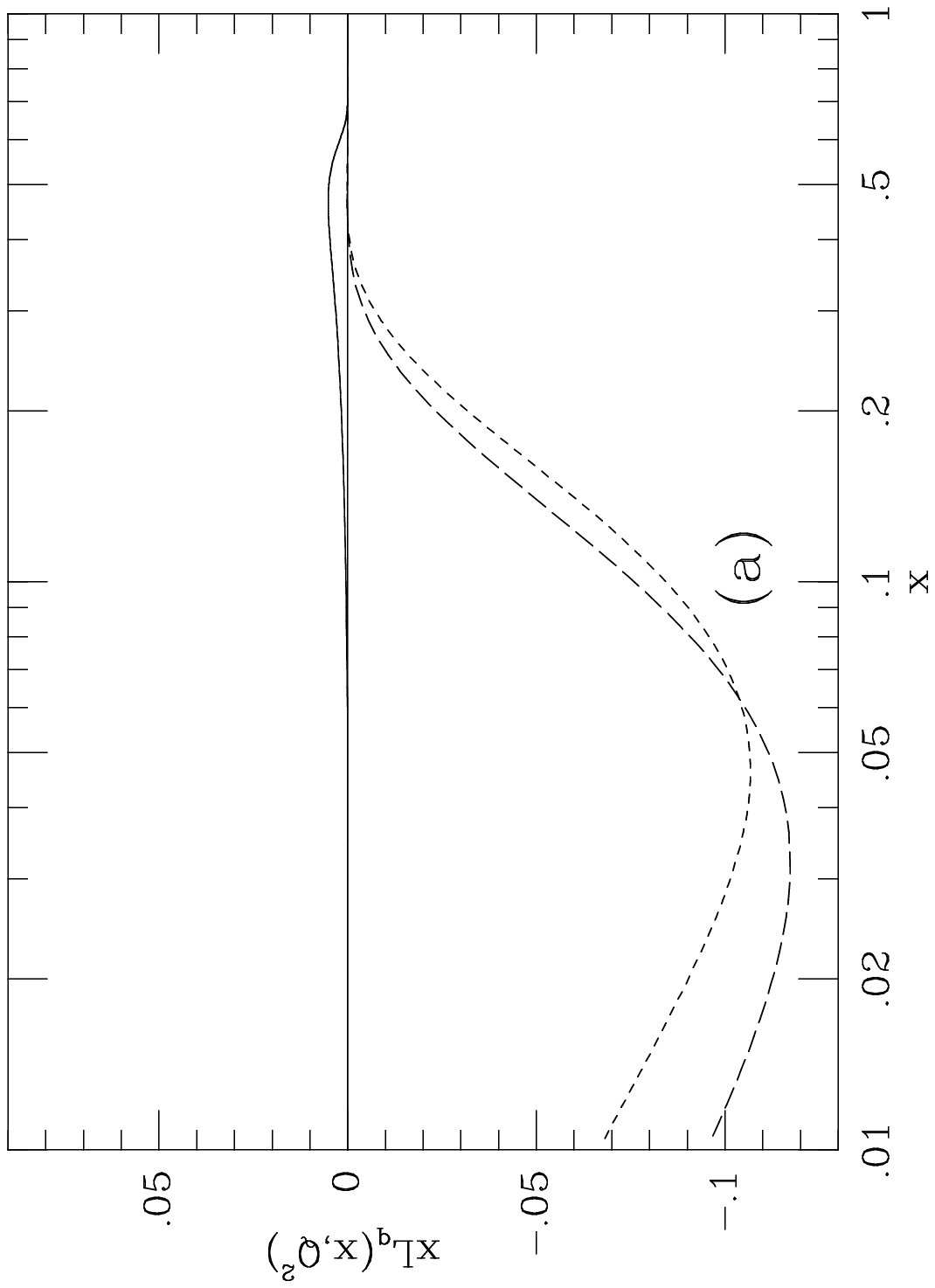}
\includegraphics{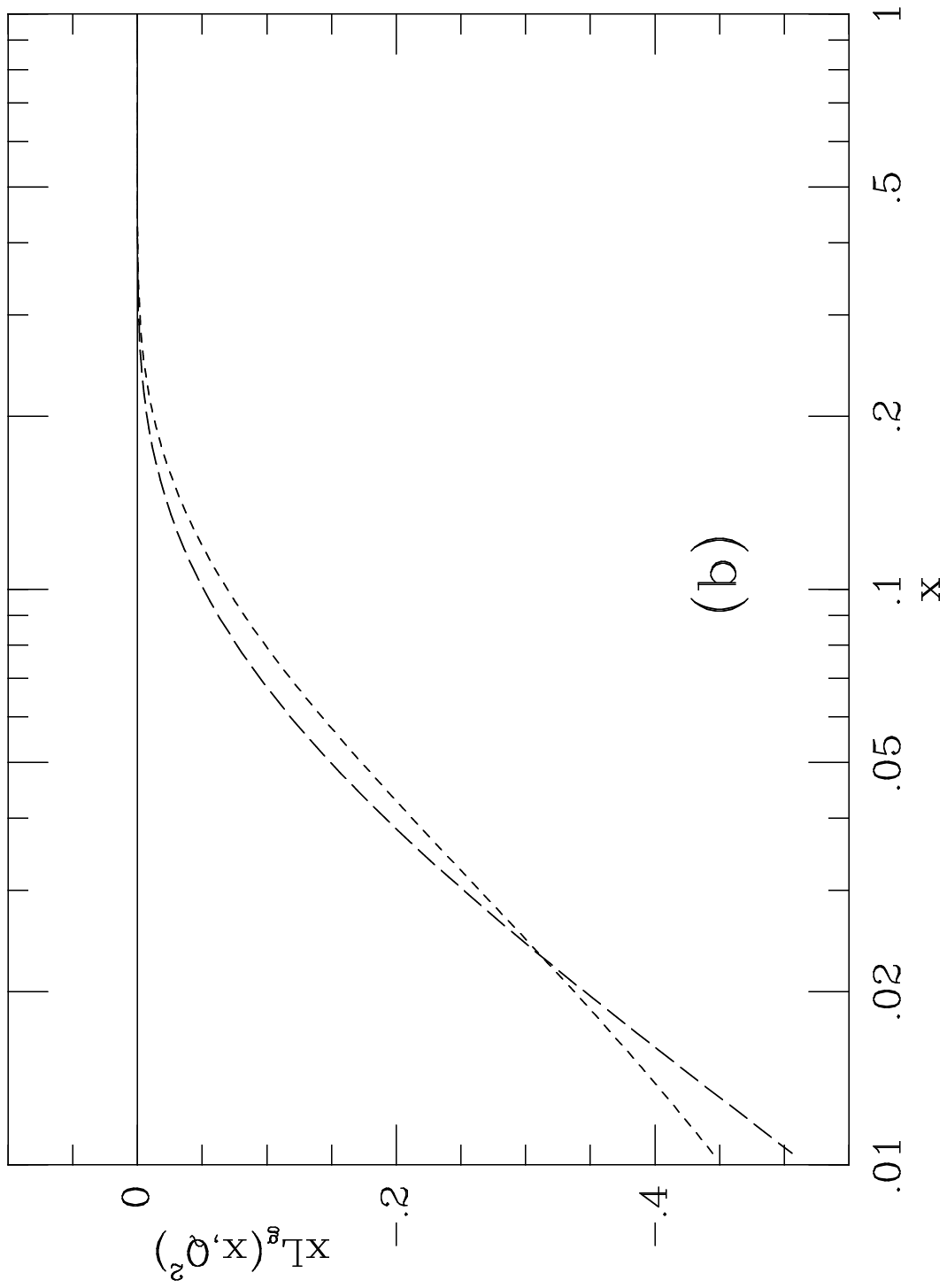}
\end{figure}
\vspace{2cm}
\centerline{\large S. Scopetta and V. Vento}
\vspace{1cm}
\centerline{\bf \large FIGURE 1}

\newpage

\begin{figure}[h]
\vspace{10cm}
\includegraphics{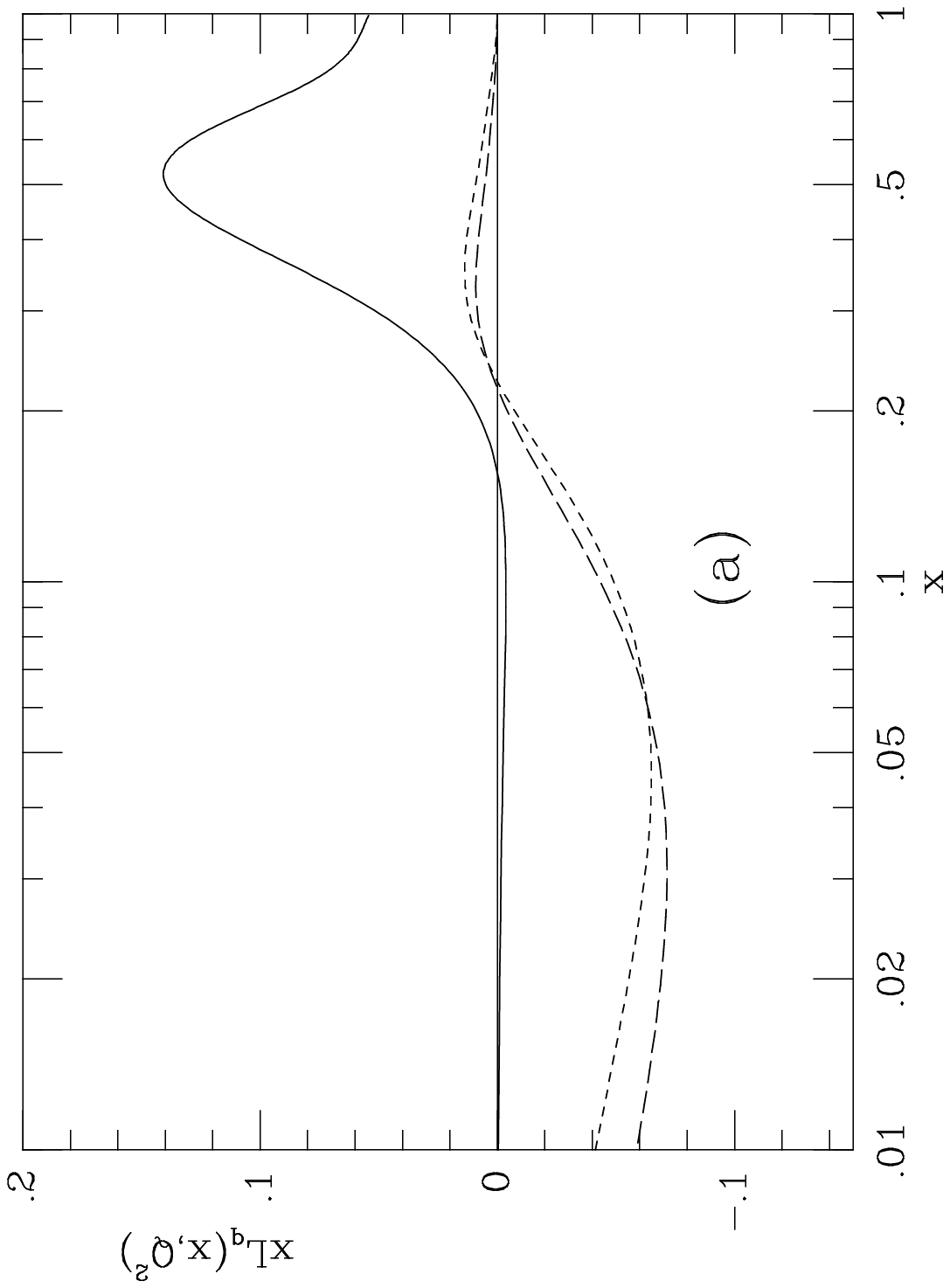}
\includegraphics{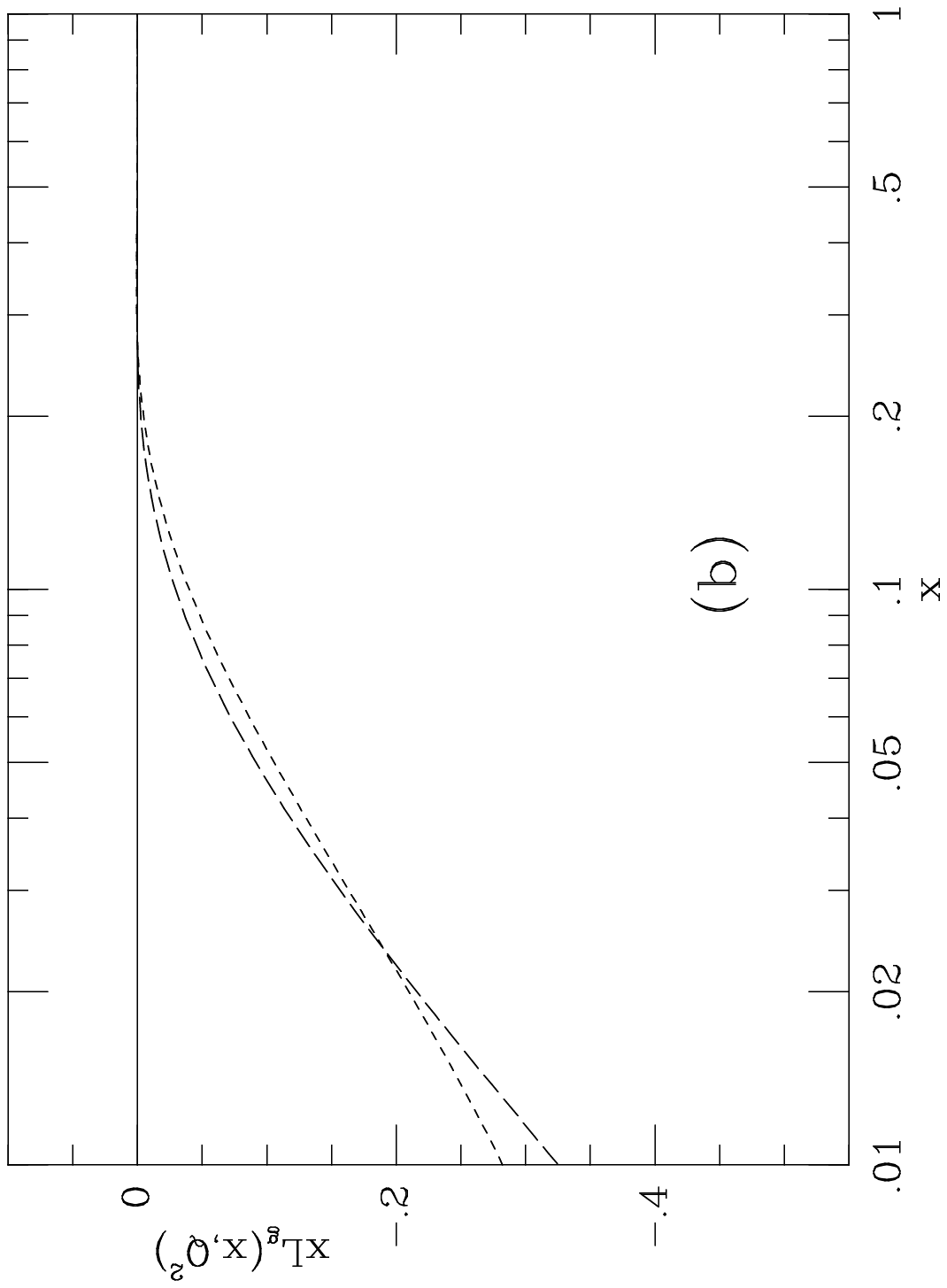}
\end{figure}
\vspace{2cm}
\centerline{\large S. Scopetta and V. Vento}
\vspace{1cm}
\centerline{\bf \large FIGURE 2}

\newpage

\begin{figure}[h]
\vspace{10cm}
\includegraphics{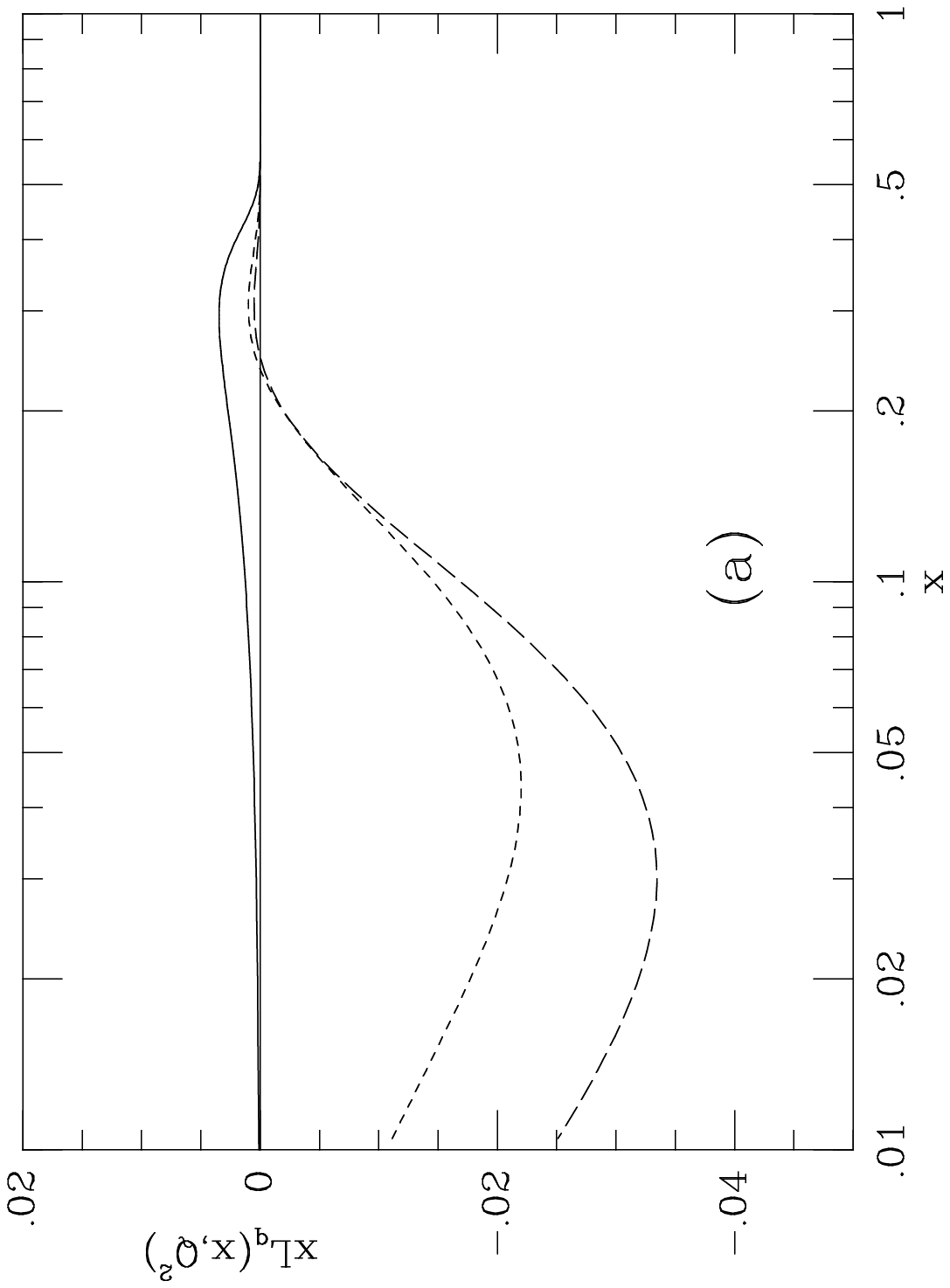}
\includegraphics{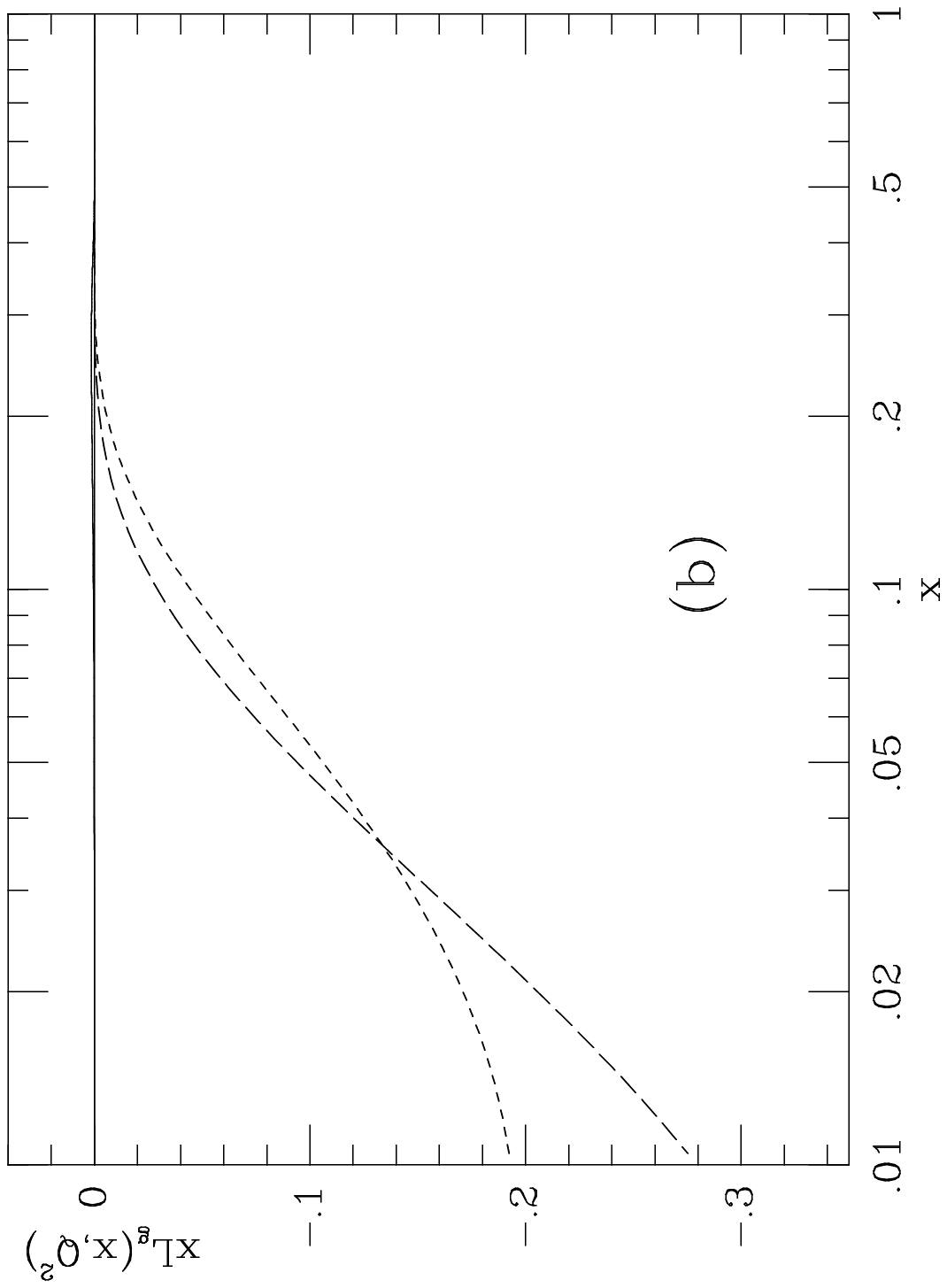}
\end{figure}
\vspace{2cm}
\centerline{\large S. Scopetta and V. Vento}
\vspace{1cm}
\centerline{\bf \large FIGURE 3}

\newpage

\begin{figure}[h]
\vspace{10cm}
\includegraphics{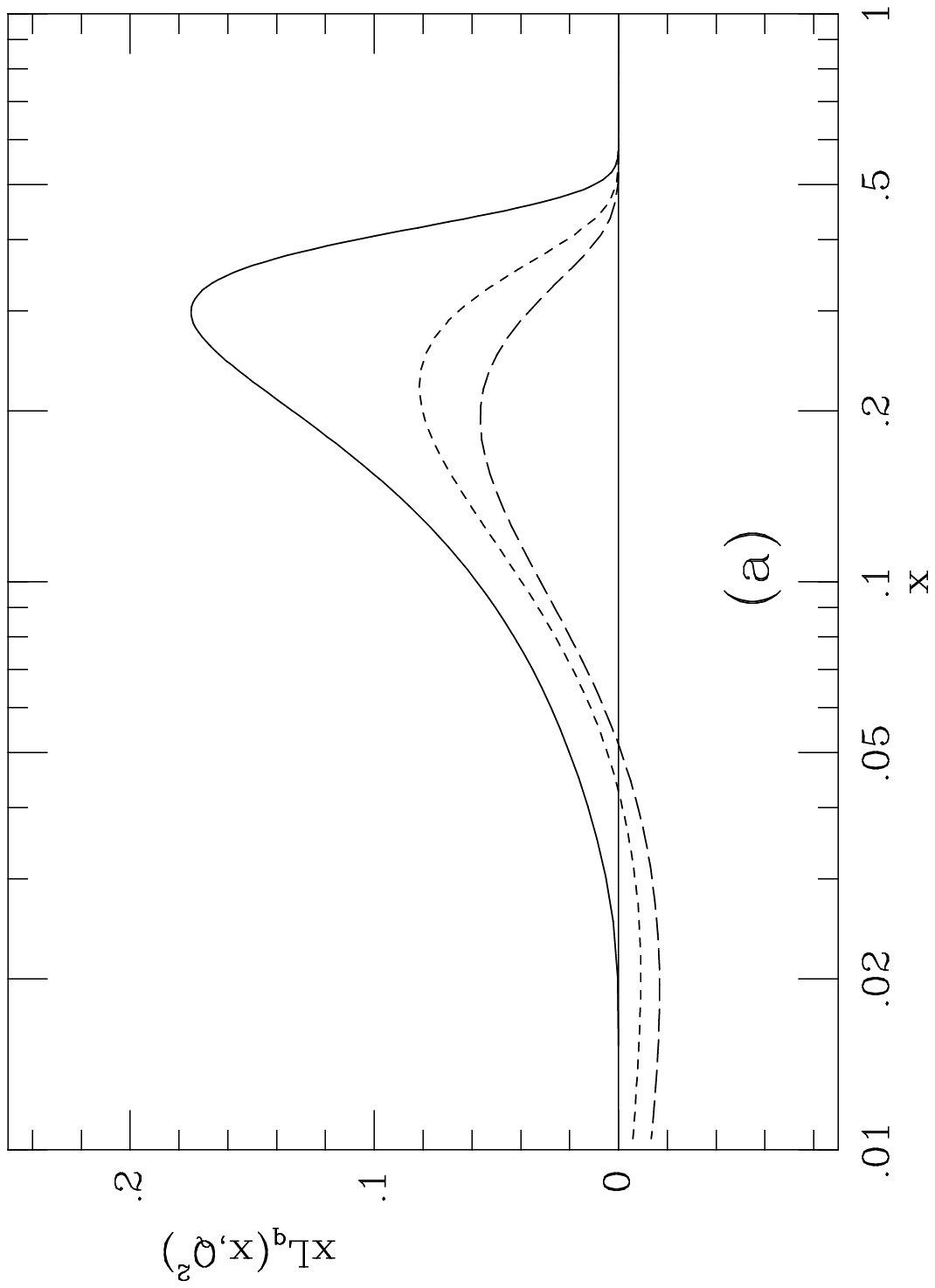}
\includegraphics{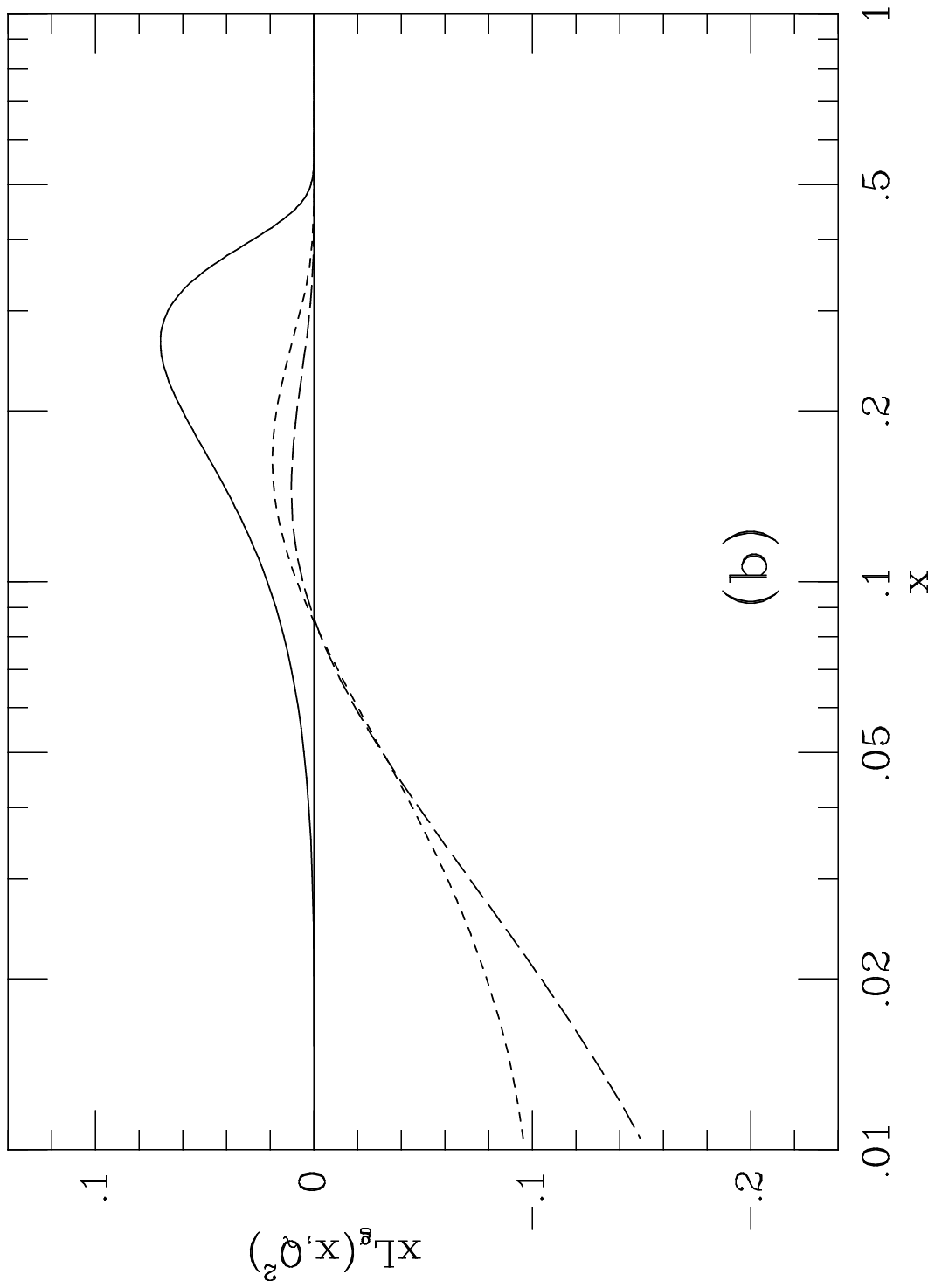}
\end{figure}
\vspace{2cm}
\centerline{\large S. Scopetta and V. Vento}
\vspace{1cm}
\centerline{\bf \large FIGURE 4}

\newpage

\begin{figure}[h]
\vspace{10cm}
\includegraphics{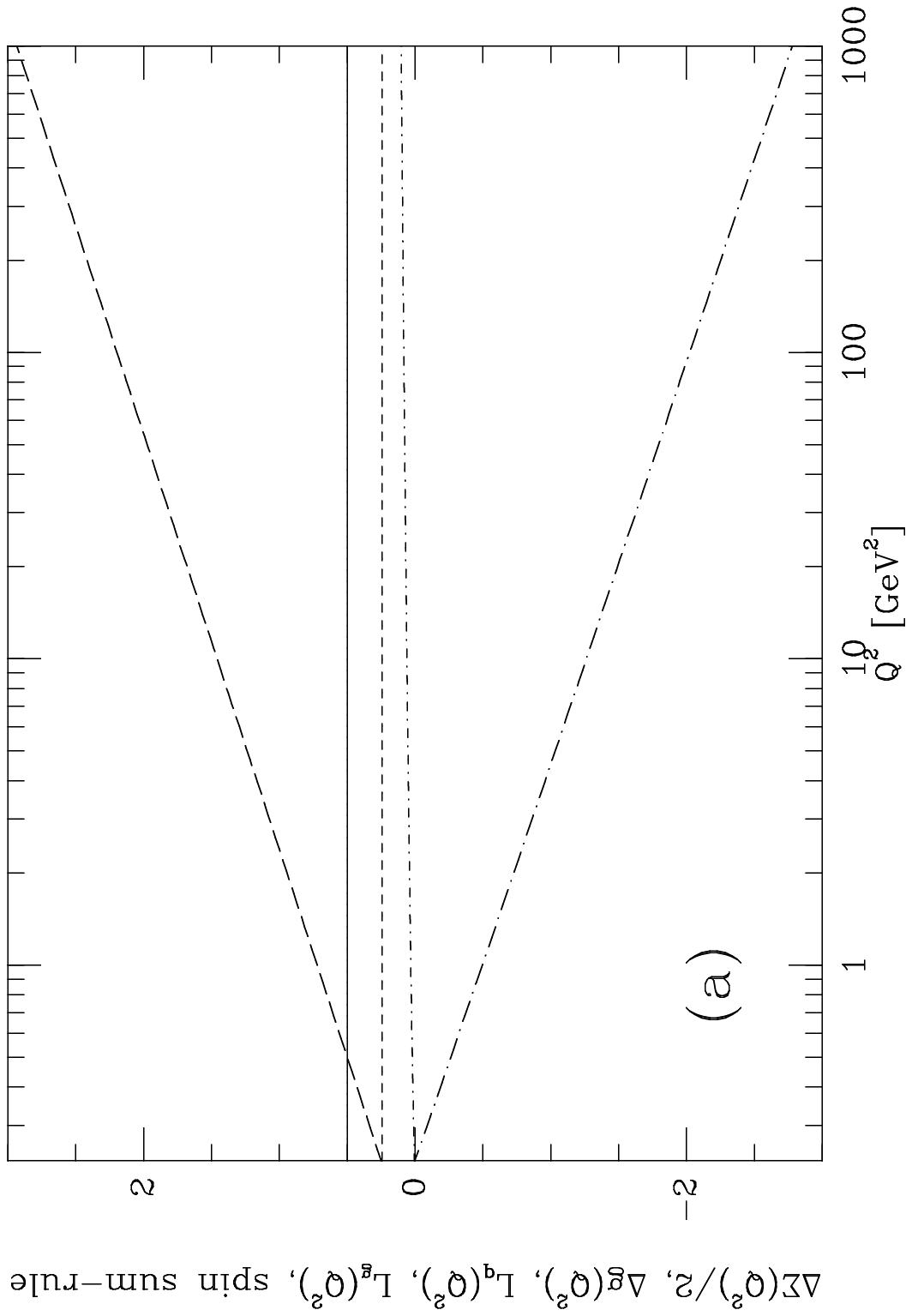}
\includegraphics{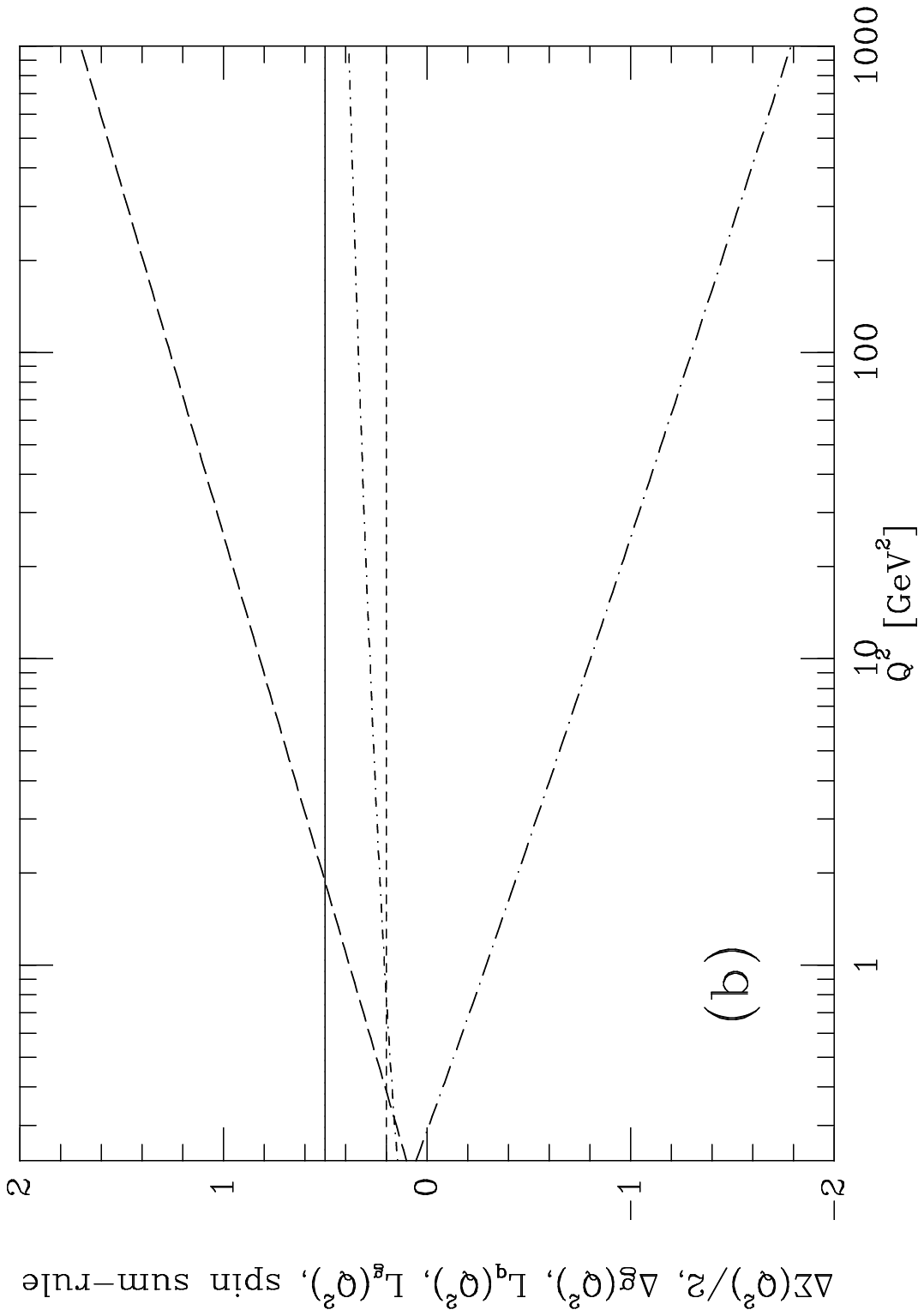}
\end{figure}
\vspace{2cm}
\centerline{\large S. Scopetta and V. Vento}
\vspace{1cm}
\centerline{\bf \large FIGURE 5}

\end{document}